# Analysis of Stick-Slip Motion as a Jump Phenomenon

Vinay A. Juvekar*, Arun K. Singh**,
*Department of Chemical Engineering, Indian Institute of Technology Bombay, Mumbai- 400076, India
**Department of Mechanical Engineering, Visvesvaraya National Institute of Technology, Nagpur- 440010, India
E-mail: vinay.juvekar@gmail.com and aksinghb@gmail.com

**Abstract**

In this work, we analyse the stick-slip motion of a soft elastomeric block on a smooth, hard surface under the application of shear, which is induced by a puller moving at a steady velocity. The frictional stress is generated by make-break of bonds between the pendent chains of the elastomeric block and bonding sites on the hard surface. Relation between velocity and frictional stress has been estimated using the bond-population balance model. Stick-slip motion occurs when the pulling velocity is lower than a critical value. Unlike, the rate-and-state friction model which views the stick-slip motion as a limit cycle, we show that during the stick phase, the sliding surface actually sticks to the hard surface and remains stationary till the shear exerted by puller causes rupture of all bonds between contacting surfaces. The major fraction of the bonds undergo catastrophic rupture so as to cause the sliding surface to slip and attain a significantly higher velocity than the pulling velocity. During the slip phase, the sliding friction is balanced by rapid make-break of weak bonds. As the sliding velocity decreases, the bonds undergo aging and the adhesion stress increases. When the bond adhesion stress exceeds the pulling stress, the contacting surfaces stick together. We have mathematically modeled both the stick and the slip regimes using the bond-population balance model. We have validated the model using the experimental data from the work of Baumberger et al (2002) on sliding of an elastomeric gelatine-gel block on a glass surface.

1. Introduction

Sliding motion of a soft solid on a hard surface is encountered in many natural phenomena as well as manmade processes/devices. A few examples include motion of gecko on a wall (McGuiggan et al, 2006), peeling of an adhesive tape from a hard surface (Maguis and Barquins, 1988), rolling of rubber tires on road (Persson, 2000), action of breaks on wheel surface (Choi et al., 2023), sliding of windshield wipers (Theodore-et-al-1992), motion of prosthetic devices (Lee et al., 2015), movement of soft robots (Chen et al., 2023) etc. In most of these cases, the soft materials are elastomers having a large number of pendent polymeric chains emanating from their surfaces. Bonding and debonding of these chains with the hard surface is the source of friction between the sliding surfaces (Persson, 2013; Ghatak et al., 2000).

A mechanical instability which is often associated sliding motion is the stick-slip instability ( Berman et al., 1996; Filippov et al., 2004;Yamaguchi et al., 2009; Corbi et al., 2012, Tian et al., 2018). The stick-slip is manifested as oscillatory motion in contrast to steady motion under the normal circumstances. Some examples of the stick-slip motion involving soft solids include squeaking of brakes (Choi et al., 2023), pulsating motion of spiny lobsters (Patek and Baio, 2007), squawking of violin bow string (Popp and Stelter, 1990), articular joints in human body (Lee et al., 2015).

Stick-slip phenomenon is under active research beginning from the work of Lord Rayleigh (see Kosterin and Kragel'skii,1960). Brockley and Ko (1970) showed that stick-slip is observed when the friction-force decreases with increase in sliding velocity. This regime of motion is called velocity weakening regime (Persson, 2013). In their model, the inertia of the sliding solid plays a dominant role. Since, in most cases, soft solids have very small mass, the inertial effect is negligibly small and cannot be invoked in order to explain the stick-slip phenomenon. Ruina (1983) proposed the rate and state friction (RSF) model in which a velocity-and-time dependent state parameter $\theta$ is included. Increase in the value of $\theta$ increases the friction stress. The value of $\theta$ increases with time at low sliding velocity, but the time dependence of $\theta$ decreases with increase in the sliding velocity and at sufficiently high sliding velocity, $\theta$ decreases with time. Using this model, Ruina (1983) was able explain the stick-slip phenomenon without invoking the inertia of the sliding solid. It is also postulated that, for a given sliding velocity, there is critical stiffness of the sliding solid such that the stick-slip is observed only when the actual stiffness is lower than the critical value. The necessary condition for occurrence of stick-slip is that the sliding is in the velocity weakening regime. Moreover, the stick-slip motion is viewed as a limit cycle (Ruina, 1983).

The present model differs from the RSF model in a number of ways. We view the stick phase as one when the sliding surface is stuck to the hard surface ( i.e. having zero sliding velocity).

The second difference is that transition from the stick to the slip phase occurs almost instantaneously. Thus the surface remains stuck till the stress generated by deformation of the soft solid exceeds a limit at which the surface velocity jumps to the slip velocity which is much greater than the pulling velocity. The reverse jump in velocity occurs during the transition from the slip to stick regime. The third difference is that during the slip phase, the mode of motion follows velocity strengthening regime ( i.e., where the frictional force increases with increase in velocity). The present model is an extension of our previous work (Singh et al., 2021), The model is validated using the experimental data of Baumberger et al. (2002).

## 2. Model Development

The basic model is described in detail in our previous works (Singh and Juvekar, 2011, Singh et al. 2021). It is a modified form of the population balance of bonds which was originally proposed by Schallamach (1963). We briefly discuss model before extending it to the stick-slip process. We consider an elastomeric block having rectangular cross-section and height $h$, placed on a plane, smooth and hard surface. Without losing generality, we can assume that the block has a unit area of contact with the hard surface. Sliding of the block on the surface is induced by pulling the upper face of the block at a constant velocity $V_0$ in the direction parallel to the base. When the upper surface of the block is pulled, block undergoes shear deformation, and exerts shear stress on the interface. The pendent chains at the base of the block form bonds with the sites on the hard surface. The interfacial shear stress causes the bonded chains to extend. Under steady sliding condition, the tension produced in these chains balances the shear stress . The bonds formed between the chains and the surface sites break both by thermal activation and due to tension in the attached chains. Detached chains can form bonds again but at different locations on the hard surface as the block slides. Bonds are thus formed, undergo aging on the hard surface and break. At any time, there is a population of bonds with different ages. The age of a bond is associated with the extent of stretching of the chain forming the bond. The population of bonds is characterized by number density $n(t_a, t)$, written as a function of the age $t_a$ of the bond and the real time $t$

$$\frac{\partial n(t_a,t)}{\partial t} = -\frac{\partial n(t_a,t)}{\partial t_a} + \frac{1}{\tau}[N_0 - N(t)]\delta(t_a) - \frac{u_0}{\tau}\exp\left(\frac{\lambda f(t_a,t)}{kT} - \frac{t_a}{\tau_a}e^{-\frac{V(t)}{V_a}}\right)n(t_a,t) \qquad (1)$$

The term to the left represents the rate of variation of number density of bonds with time, whereas the first term to the right represents its rate of variation in the age space. The second term to the right represents the rate of birth of the bonds at time $t$ and is proportional to the number of vacant sites on the surface, which is the difference between $N_0$, the total number of the occupiable sites and $N(t)$, the number of the bonded sites at time $t$ . The term $\tau$ is the characteristic time constant associated with the thermal make-break process (Schallamach, 1963). Since new-born bonds have zero age, the population peaks at $t_a = 0$ as indicated by inclusion of the Dirac delta function $\delta(t_a)$. The number of the bonded sites at time $t$ ,i.e. $N(t)$, is obtained by integrating $n(t_a, t)$ over the age space as shown in Eq 2 below. Here, $t_m$ is the upper limit of the age of a bond and is assumed to be the same for all bonds

$$N(t) = \int_0^{t_m} n(t_a, t) dt_a \qquad (2)$$

The third term to the right of Eq 1 represents the rate of death (breakage) of bonds. The term $u_0 = \exp(-W_0/kT)$ is the bond breakage probability weight for newly formed bonds i.e. with $t_a = 0$, $W_0$ being the corresponding bond adhesion energy. The exponential term accompanying $u_0$ includes two corrections. The term $\exp[\lambda f(t_a, t)/kT]$ represents the increase in the probability of breakage of a bond due to increase in the activation energy associated with stretching of the chain attached to the bond (Eyring factor). In this term, $\lambda$ is the activation length and $f(t_a)$ is the force exerted on the bond by the attached chain due to its extension. The latter is related to the sliding velocity $V(t)$ by the finite extensive nonlinear elastic (FENE) model

$$f(t_a, t) = \frac{M t_a V(t)}{1 - [t_a V(t)/\Delta l_m]^2} \qquad (3)$$

In this equation $M$ is the chain stiffness and $\Delta l_m$ is maximum attainable extension of the chain. This extension sets the upper limit on the age of a bond in Eq 2 as $t_m = \Delta l_m/V(t)$, since $f(t_m, t) = \infty$ .

The term $\frac{t_a}{\tau_a} \exp[-V(t)/V_a]$ in Eq 1 accounts for the velocity dependent aging of the bonds (Singh et al., 2021), $\tau_a$ being the aging time constant and $V_a$ is the characteristic velocity associated with bond aging. When sliding velocity is much greater than $V_a$, the aging term becomes zero and bonds do not strengthen with age. Only if $V(t)$ is comparable or less than $V_a$, bonds age and the rate of breakage decreases with increase in bond age $t_a$.

The frictional stress $\sigma(t)$ at any time $t$ is given by summing the forces experienced by all bonds

$$\sigma(t) = MV(t) \int_0^{\Delta l_m/V(t)} n(t_a, t) \frac{t_a}{1 - [t_a V(t)/\Delta l_m]^2} dt_a \qquad (4)$$

Under unsteady sliding motion, there exists a difference between the pulling velocity $V_0$ and the sliding velocity $V(t)$ causing variation in the shear deformation of the block with time. If we assume that the sliding is occurring under quasistatic condition, then the stress generated by this deformation must balance the friction stress. Also, the rate of change of two stresses must also match. This balance is give by

$$\frac{d\sigma(t)}{dt} = \frac{G_d}{h}[V_0 - V(t)] \tag{5}$$

where, $G_d$ is the shear modulus of the elastomer ($G_d/h$ represents the shear stiffness of the block)

When the block is sliding at steady state, the time derivative of frictional stress in Eq 5 is zero. This gives $V(t) = V_0$, i.e. the sliding velocity is equal to pulling velocity. Eq 1 then reduces to

$$\frac{dn(t_a)}{dt_a} = -\frac{u_0}{\tau}\exp\left(\frac{\lambda}{kT}\frac{Mt_aV_0}{1-[t_aV_0/\Delta l_m]^2} - \frac{t_a}{\tau_a}e^{-\frac{V_0}{V_a}}\right)n(t_a) \tag{6}$$

The corresponding initial condition is

$$n(t_a = 0) = \frac{1}{\tau}\left[N_0 - \int_0^{\Delta l_m/V_0} n(t_a)dt_a\right] \tag{7}$$

Moreover, Eq 4 reduces to

$$\sigma = MV_0 \int_0^{\Delta l_m/V_0} \frac{t_a n(t_a)}{1-[t_aV_0/\Delta l_m]^2} dt_a \tag{8}$$

A typical plot of $\sigma$ against $V_0$, as predicted by Eq 6 to 8, is shown in Figure 1.

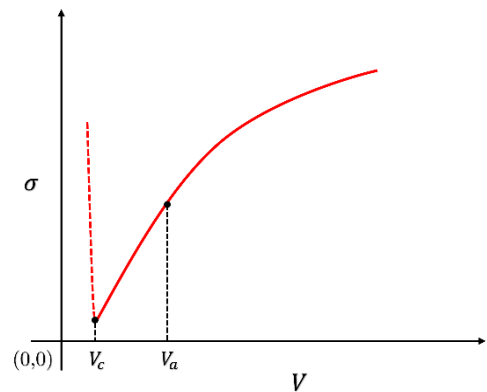

**Figure 1: Plot of friction stress versus sliding velocity as predicted by Eq 6 to 8.**

The plot has a minimum at $V_0 = V_c$ such that

$$\left(\frac{d\sigma}{dV_0}\right)_{V_c} = 0 \tag{9}$$

This point divides the curve in two branches. On the right branch, stress increases with increase in velocity (velocity strengthening region), whereas on the left branch, stress decreases with increase in velocity ( velocity weakening region). Point $V_a$, on the right branch divides it further into two regions. In the region where $V_0 \gg V_a$, we have $\exp(-V_0/V_a) \to 0$ and the term associated with bond strengthening in Eq 6, can be neglected. In this region of high velocity, chains are pulled rapidly and are not allowed sufficient time to age on the surface. As a result, they form weak bonds with the surface. On the other hand in the region where $V_0$ is comparable to or less than $V_a$, bonds are allowed sufficient time to age and become strong. Higher bond strength reduces the rate of breakage of bonds and the friction stress decreases more slowly with decrease in the velocity than in the absence of aging. At $V_0 = V_c$ the rate of change of frictional stress with velocity becomes zero and beyond this point, increase in the bond strength due to aging is so high that the frictional stress increases with decreasing velocity.

We now extend this analysis to unsteady sliding. For weak bonds, the time scale of the make-and-break of bonds is of the order of microseconds (Vorvolakos and Chaudhury, 2003). For soft solids, pulling stress is comparable with the shear modulus $G_d$ of the solid and hence from Eq 5, the time scale of change in pulling stress is of the order of $h/V_0$ and since $h$ is of the order of millimetres, and the pulling velocity is of the order of mm.s$^{-1}$, the time scale of change in pulling stress is of the order of seconds. This means the process of make-break of bonds is very rapid and attains steady state in much shorter time compared to time duration of change in the imposed shear. Hence make-break of bonds can be viewed as a quasistatic process. In this case, Eq 1 reduces to

$$\frac{dn(t_a,t)}{dt_a} = -\frac{u_0}{\tau}\exp\left(\frac{\lambda}{kT}\frac{M t_a V(t)}{1-[t_a V(t)/\Delta l_m]^2} - \frac{t_a}{\tau_a}e^{-\frac{V(t)}{V_a}}\right)n(t_a,t) \tag{9}$$

Time dependence in Eq 9 arises due to $V(t)$, the sliding velocity which depends on time as described by Eq 5. Figure 1 still represents the stress-velocity relation except for the fact that, the abscissa $V_0$ ( the pulling velocity) should be replaced by $V(t)$ the sliding velocity.

Now consider the case where the block is sliding at a steady velocity V(0), lying on the right branch of Figure 1. Suppose at some reference time $t = 0$, pulling velocity is stepped down. We denote the pulling velocity after the step change as $V_0$. Note that V(0) corresponds to the sliding velocity before the step change and constitutes the initial condition for the sliding velocity in Eq 5.

We first consider the case where, $V_0$ lies on the right branch of Figure 1. Since $V(0) > V_0$, the shear stress $d\sigma/dt$ is negative as per Eq 5 and hence the frictional stress decreases with time. This causes velocity to decrease with time along the curve in Figure 1, until it reaches $V_0$, which is the new steady state sliding velocity. The rate of change of velocity during this period is governed by the following equation, which can be derived from Eq 5

$$\frac{dV(t)}{dt} = -\frac{G_d}{h}\left[\frac{V(t)-V_0}{d\sigma(t)/dV(t)}\right] \qquad (10)$$

In this equation, $d\sigma(t)/dV(t)$ can be obtained in terms of $V(t)$ by differentiation of $\sigma(t)$, obtained from the solution of Eq 4, with respect to $V(t)$. The initial condition for Eq 10 is $V(t) = V(0)$.

Next, we consider the case when the velocity after the step change, i.e. $V_0$, lies on the left branch of Figure 1. When the sliding velocity $V(t)$ crosses $V_a$, the bonds begin to age and become stronger. Next, when the sliding velocity reaches $V_c$, where $d\sigma/dV(t) = 0$, we see from Eq 10, that $dV(t)/dt = -\infty$. This causes rapid deceleration of the sliding velocity. Moreover, when $V(t)$ falls below $V_c$, equality given by Eq 5 is violated. The reason is that the difference $V_0 - V(t)$ is negative and hence according to Eq 5, the pulling stress decreases with time. However, as velocity decreases along the stress-velocity curve, the adhesive stress increases with time. As a result the pulling stress is not able to overcome the adhesive stress when the pulling velocity falls below $V_c$. The quasi-static condition is violated and stress-velocity relation schematically follows path 2 in Figure 2.

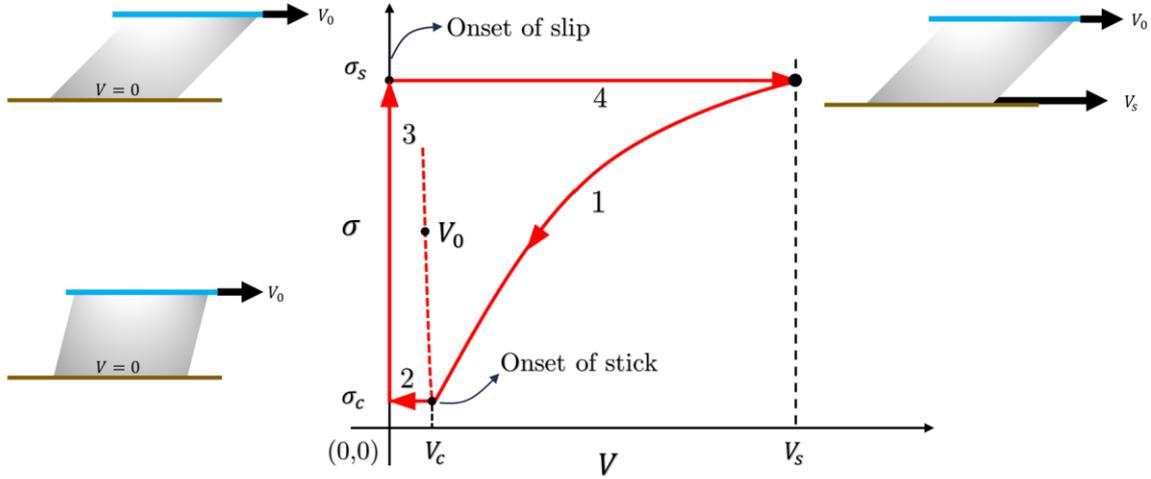

**Fig 2. Representation of the stick-slip process on the stress-velocity diagram.**

This causes a rapid aging of bonds, thereby increasing the sticking tendency. When $V(t)$ falls below $V_0$, the pulling stress begins increase with time. However, the sticking stress increases even more rapidly and sliding velocity continues to decrease till it becomes zero. Therefore, in practice, the left branch of the quasistatic profile is never realised. Since transition of the sliding velocity from $V_c$ to zero happens rapidly, we can view this process as a jump in the sliding velocity.

Once stuck, the sliding surface remains at zero velocity, and hence the stress velocity relation follows path 3 in Figure 2. During this period, the top surface of the block move with velocity $V_0$. As a result, the block is subjected to an increasing shear deformation with time. Initially, the bonds are subjected to lower stress, which allows them to undergo further aging. Since the surface in contact is stationary, all bonds age together. Hence they all have nearly the same strength. Since the surface is stationary, the extent of breakage of the bonds is small. When the shear stress becomes sufficiently high, rapid breakup of bonds occurs as explained later. The stress at which complete breakage of bonds occurs is denoted by $\sigma_s$.

When the stress $\sigma_s$ is reached, the sliding surface slips and its velocity raises rapidly until the new bonds are formed. The new bonds are weaker bonds (i.e. those bonds which do not age) and much higher sliding velocity $V_s$ is needed in order to balance the shear stress. Since $V_s > V_0$, sliding velocity decreases with time along path 1 in Figure 2. Paths 1, 2, 3, 4, therefore, generate a stick-slip cycle along which the stress and the velocity varies.

During the stick phase, the sliding surface is at zero velocity. Hence Equation 5 reduces to

$$\frac{d\sigma(t)}{dt} = \frac{G_d}{h} V_0 \tag{11}$$

Eq 11 on integration with initial condition $\sigma(0) = \sigma_c$ yields the following equation.

$$\sigma(t) = \frac{G_d}{h} V_0 t + \sigma_c \tag{12}$$

The most important characteristic of the stick phase is that during this phase, the stress varies linearly with time. This is evident from Eq 12.

Since the time allowed for the chains to bond to the hard surface is sufficiently long, each chain bonds with multiple sites on the surface. Hence, the number of chains attached to the surface is much less than those during the slip phase, when each chain occupies a single site. It is therefore convenient to replace the population balance of bonds by population balance of the attached chains. We assume that all attached chains have equal age at any time. The chains which are initially attached, undergo detachment with time when they undergo extension under the application of the imposed stress. Although new chains are likely to be attached in the place of the detached chains, we expect their contribution to load bearing is small for the following reason. During the initial period of the stick phase, chain undergo extension with a low probability of detachment. The new chains are attached only near the later part of the stick phase, where the rapid detachment of the previously attached chains occurs. Hence, the newly attached chains do not have sufficient time to age on the surface and their extension is much smaller compared to the chains which were attached in the beginning. Therefore, we neglect the contribution to stress from the newly attached chains. Hence we assume that the age of all attached chains is the same as the time elapsed during the stick phase. Consequently, the population balance equation for the attached chains can be written in the time domain rather than the age domain. Moreover, the population balance equation has only the death and the aging terms. It can be therefore written as

$$\frac{dN_c(t)}{dt} = -\frac{u_c}{\tau} \exp\left(\frac{\lambda f_c(t)}{kT} - \frac{t}{\tau_a}\right) N_c(t) \tag{13}$$

In this equation, time is measured from the instant of stick. $N_c(t)$ is the number of attached chains at time $t$ during the stick phase, $t = 0$ corresponds to the beginning of the stick phase, at which the number of the attached chains is $N_{c0}$. The term $u_c$ is the bond breakage probability weight at the beginning of the stick phase. This is expected to be much lower than $u_0$ since significant aging of bonds occur before the sliding surface comes to standstill. The term $f_c(t)$ is the force acting on a single chain due to shear exerted by the deformation of the block. We assume that all chains equally share this force. Hence

$$f_c(t) = \frac{\sigma(t)}{N_c(t)} \tag{14}$$

where, $\sigma(t)$ is given by Eq 12.

Combining Eq 12 to 14, we obtain

$$\frac{dN_c(t)}{dt} = -\frac{u_c}{\tau} \exp\left\{\frac{\lambda}{kT}\frac{1}{N_c(t)}\left(\frac{G_d}{h}V_0 t + \sigma_c\right) - \frac{t}{\tau_a}\right\} N_c(t) \tag{15}$$

Eq 15 is an ordinary differential equation to be solved using the initial condition $N_c(0) = N_{c0}$. The time $t_s$ is time at which $N_c$ reduces to zero, i.e. $N_c(t_s) = 0$, The corresponding stress is

$$\sigma_s = \frac{G_d}{h}V_0 t_s + \sigma_c \tag{16}$$

A typical plot of $N_c/N_{c0}$ against time is shown in Figure 3. It is seen that during the initial period, the rate of decrees of $N_c/N_{c0}$ with time is very slow. The reason is that all attached chains equally share the pulling stress (notice that $N_c(t)$ in the denominator of the first term in exponential of Eq 15). Since initially, the number of attached chains is large, load on each chain is small and hence probability of its detachment is small. Once the chains begin to detach, $N_c(t)$ decreses and since it is in the denominator of the exponential terms, probability of detachment of chains is amplified. This causes a rapid fall in $N_c/N_{c0}$ with time as seen in the Figure 3.

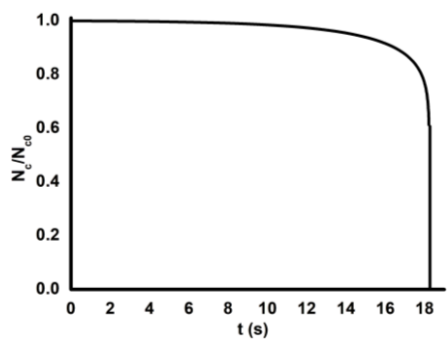

**Figure 3: Typical plot of $N_c/N_{c0}$ against time as obtained through the solution of Eq 15.**

### 3. Model Validation

We validate the model developed in the previous section for explaining the stick-slip process. For this purpose, we have used the experimental data from the paper by Baumberger et al.

(2002). These data pertain to sliding of a gelatine gel block on a glass surface. The gel samples were prepared from 5 wt % solution of gelatine ( bloom strength =300) in water. Experiments were conducted at 20°C. Shear modulus of the gel, $G_d$ was $4\ kPa$. The height of the gel block, $h$, was $10mm$. The stiffness of the block is $\frac{G_d}{h} = 400\ Pa.m^{-1}$. Since the sliding block has trapezoidal shape, Eq 5 needs to be modified to the following form

$$\frac{d\sigma(t)}{dt} = \frac{G_d}{h}\left[\frac{\ln(B/l)}{(B/l)-1}\right][V_0 - V(t)] \tag{17}$$

Here, $B = 50$ mm is the length of the trapezoid in contact with the puller and $l = 30\ mm$ is length of the trapezoid in contact with the glass plate. Substitution of these lengths in Eq 15 gives the geometric correction as 0.766.

The experiments have been performed both under the steady sliding condition and the stick-slip condition. We first fit the experimental data for steady sliding. Experimental data are presented as shear stress versus sliding velocity at discrete points in the range of velocity of $140 - 2000\ \mu m.s^{-1}$. These data are found in Figure 3 from the paper by Baumberger et al. (2002) and are plotted in Figure 4 below as points. We fit Eq 6 to 8 to these data and obtain the model parameters. We perform the regression in two steps. In the velocity range: $800 - 2000\ \mu m.s^{-1}$, we neglect the velocity dependent aging of bonds ( we assume that the $V_a \ll 800\ \mu m.s^{-1}$ so that the exponential aging term is absent in Eq 6). This reduces the model parameters to five, namely, $u_0$, $N_0$, $\tau$, $\lambda M/kT$, and $\Delta l_m$ (Singh et al., 2021). Among these, $u_0$ is chosen a priori as unity based on the fact that both gelatine molecules and glass surface adsorb water, Hence, at the beginning of adsorption, water molecules on the gelatine chains bond with those on the glass surface, rendering the bond-adhesion energy $W_0 = 0$. When the make-break process of bonds is very rapid, no time is allowed for water to be displaced from the contacting surfaces and aging of the bond does not occur. As a result, out of the five parameters, only four need to be estimated. The estimated parameters are : $N_0 = 1.22 \times 10^{14} m^{-2}, \tau = 224\ \mu s, \lambda M/kT = 2.47 \times 10^{12} m^{-2}$ and $\Delta l_m = 1.074 \mu m$. The plot of frictional stress vs sliding velocity obtained using this approximation is shown as a blue dotted line in Figure 4.

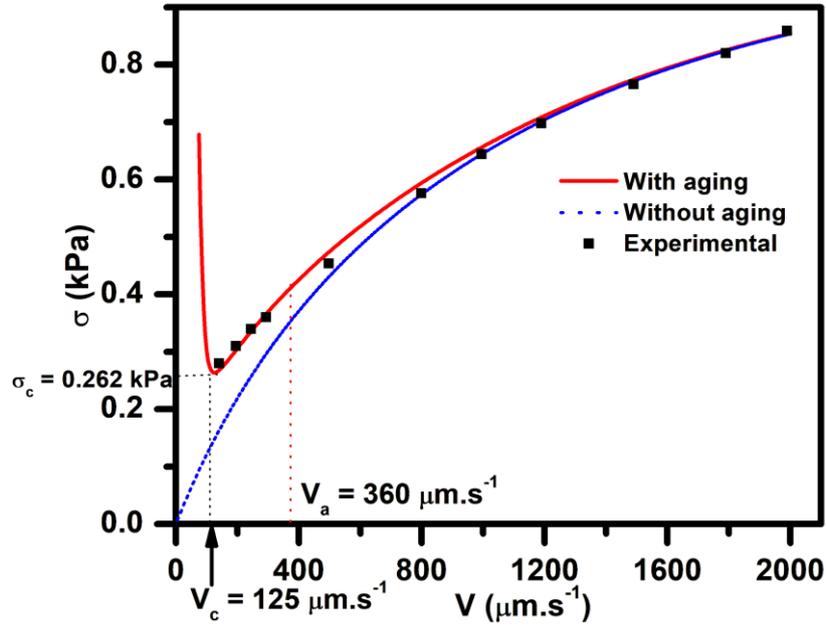

**Fig 4. Frictional stress data for steady sliding of gelatine gel on glass surface Points are the data from Figure 3 of Baumberger et al (2002), the solid curve is the model fit ( Eq 6 to 8 from this paper). The curve is extended to lower velocities to show inversion of velocity-stress relation.**

As the next step, we fit the Eq 6 to 8 including the aging term to the experimental data in the velocity range 140-800 $\mu m.s^{-1}$. Aim is to estimate the two remaining parameters, namely $\tau_A$ and $V_A$. The best fit parameters are $\tau_a = 899\ \mu s$ and $V_a = 360\ \mu m.s^{-1}$. The model fit is shown as a solid curve in Figure 4. The fit is seen to be good.

The model fit is extended beyond the experimental data to determine the critical velocity and the corresponding critical stress. The critical velocity is $V_c = 125\ \mu m.s^{-1}$ and the corresponding critical stress $\sigma_c = 259\ Pa$. It is also important to notice that the frictional stress rises very steeply with decrease in velocity beyond the minimum.

Next we analyse the experimental data in the stick-slip regime. These data are extracted from Figure 2 of Baumberger et al. (2002). The data in Figure 2(a) corresponds to pulling velocity $V_0 = 50\ \mu m.s^{-1}$, Figure 2 (b) corresponds to pulling velocity $V_0 = 100\ \mu m.s^{-1}$ and Figure 2(c) corresponds to pulling velocity $V_0 = 350\ \mu m.s^{-1}$ where the sliding reaches a steady frictional stress of $0.4\ kPa$ (Note that the pulling velocity mentioned in the description of Figure 2(c) of Baumberger et al. is $150\ \mu m.s^{-1}$, which is incorrect. The correct value of the pulling velocity is $350\ \mu m.s^{-1}$ since it is consistent with the steady frictional stress of $0.4\ kPa$ as seen from steady state data from Figure 3 of their paper and Figure 4 of the present paper).

The data extracted from the Figure 2 of Baumberger et al. (2002) are plotted as points in Figure 4 (a), 4(b) and 4(c). The solid lines are the model fits which are discussed below.

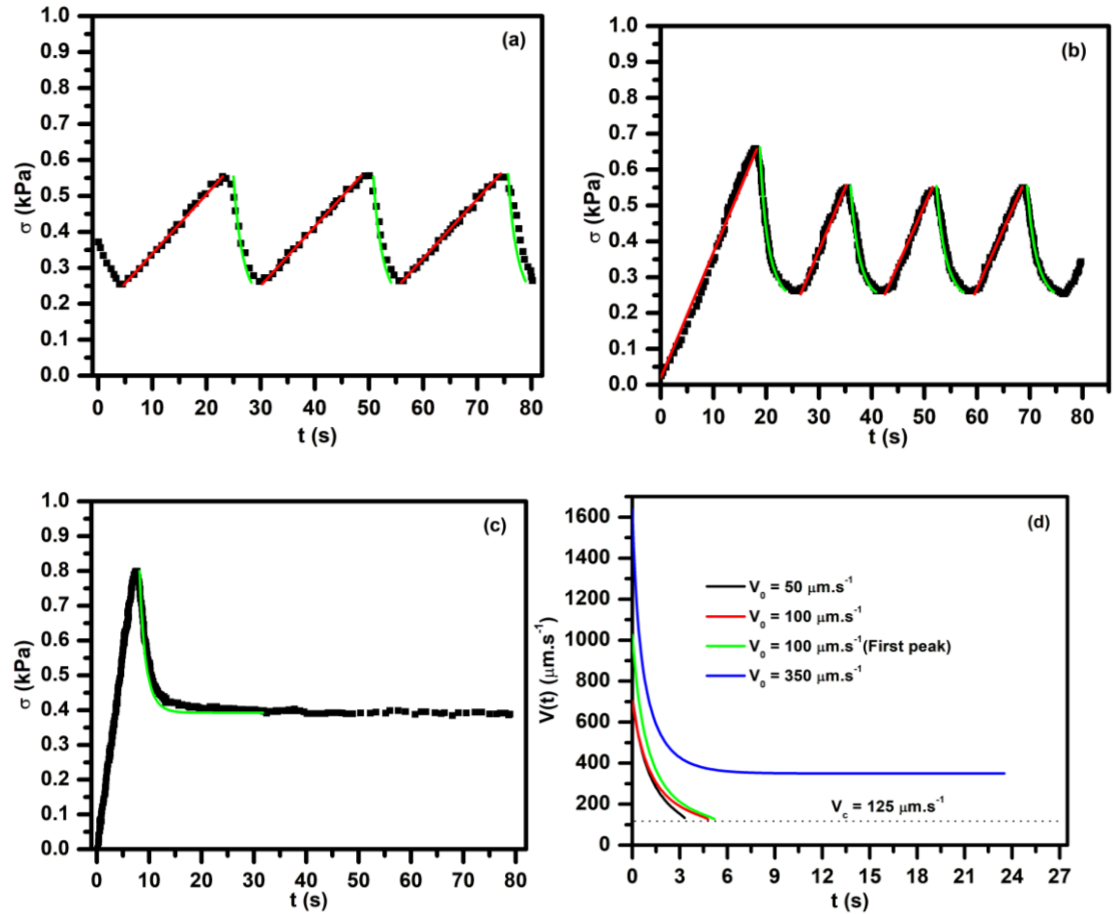

**Figure 5: (a), (b), (c) show match between the experimental data of Baumberger et al.(2002) and the present model (a) sliding velocity $50 \mu m.s^{-1}$ (b) sliding velocity 100 $\mu m.s^{-1}$ (c) sliding velocity 350 $\mu m.s^{-1}$ (d) shows variation of sliding velocity with during the slip phase as predicted by the model.**

Before proceeding for the model fit, we comment on certain characteristics of these stick slip plots. In Table 1 we have listed the characteristics of the plots in Figure 5. From the first row of Table 1, we see that the values of the critical stress, $\sigma_c$ are not only nearly equal for all cycles, but are also independent of the pulling velocity. The value of $\sigma_c$ lies in the range $259 \pm 5\ Pa$. This observation is consistent with the deduction from our model that the critical stress is independent of the pulling velocity.

**Table 1: Important characteristics of stick-slip plots shown in Figure 4 ( Experimental data from Figures 2(a) and 2(b) from Baumberger et al. 2002).**

|  | Pulling velocity, $V_0 = 50\ \mu m.s^{-1}$ | | | Pulling velocity, $V_0 = 100\ \mu m.s^{-1}$ | | | | Pulling velocity, $V_0 = 350\ \mu m.$ |
|---|---|---|---|---|---|---|---|---|
|  | Cycle 1 | Cycle 2 | Cycle 3 | Cycle 1 | Cycle 2 | Cycle 3 | Cycle 4 | Cycle 1 |
| $\sigma_c, Pa$ | 255 | 262 | 264 | 261 | 261 | 261 | 254 | -- |
| $\frac{d\sigma}{dt}, Pa.s^{-1}$, stick phase | 16.5 | 16.3 | 16.8 | 35.0 | 33.5 | 34.0 | 34.0 | 116 |
| $G_d/h$ $kPa.m^{-1}$ | 431 | 424 | 437 | 457 | 437 | 444 | 444 | 457 |
| $\sigma_s, Pa$ | 554 | 556 | 552 | 658 | 549 | 542 | 549 | 800 |
| $t_s, s$ | 18.3 | 18.3 | 17.8 | 18.2 | 8.82 | 8.61 | 8.83 | 7.51 |
| $V_s. \mu m.s^{-1}$ | 702 | 702 | 720 | 1026 | 702 | 702 | 702 | 760 |

We also find that the stress-time plots during the stick phase are straight lines. This observation is consistent with our deduction that the sliding surface is stationary during the stick phase. Their slopes are listed in Table 1. From Eq 12, we have $d\sigma(t)/dt = V_0 G_d/h \left[\frac{\ln(B/l)}{(B/l)-1}\right]$. Hence from the slope and the pulling velocity, we obtain $G_d/h$. These values are also listed in Table 1. We see that these values are nearly the same in all cycles and at all velocities. We also see that $G_d/h$ obtained by dividing the shear modulus by the height of the block is $400\ kPa.m^{-1}$, which compares well with the value listed in the table.

Next, we examine values of $\sigma_s$, stress at which transition from the stick to the slip phase occurs. We see that except for the first cycle of $V_0 = 100\ \mu m.s^{-1}$, which show a higher value of $\sigma_s$, rest of the values of $\sigma_s$ are the same for rest of the cycles. The first cycle of $V_0 = 100\ \mu m.s^{-1}$ is different since the contacting surfaces undergo extra aging during the first cycle. Even when, no extra aging time is allowed, aging occurs in the time gap between the instant when gelatine gel block is placed on glass surface and the instant when shearing of the block commences.

We then examine the values of $t_s$, the time interval over which the block is stuck to the hard surface. We call this the stick interval. The stick intervals are listed in fifth row of Table 1. We see that $t_s = 18.9 s$ for $V_0 = 50\ \mu m.s^{-1}$ and $t_s = 8.6\ s$ for $V_0 = 100\ \mu m.s^{-1}$. From Eq 15 we find that slip occurs when the exponential term on the right attains a sufficiently large value. If we neglect the contribution from $\sigma_c$ and $\tau_a$ we can say that the stick interval should scale inversely with the pulling velocity, if the aging time contribution is small . This is consistent with the values in Table 1. Effect of $1/\tau_a$ is to lengthen the stick time. This lengthening is expected to be more pronounced at lower velocity and hence $t_s$ is slightly longer than that obtained velocity alone.

As we have already mentioned, at the peak point of the stick phase, the sliding surface slips and slides with sufficiently high velocity so that the new bonds formed now balance the stress. Since at high velocity, bonds do not allow time to age, they are necessarily weak bonds. We can therefore estimate the sliding velocity $V_s$ using the following form of Eq 8

$$\sigma_s = MV_s \int_0^{\Delta l_m/V_s} \frac{t_a n(t_a)}{1-[t_a V_s/\Delta l_m]^2} dt_a \tag{8}$$

where $n(t_a)$ as a function of $t_a$ is obtained by solving Eq 6 , excluding the aging term , and using the boundary condition given by Eq 7. The values of the slip velocity are listed in the last row of Table 1.

Next we check whether the stress-time curve obtained during the slip phase can be predicted using our quasi-static model. We solve Eq 10, using the initial condition $V(0) = V_s$. For any given $V(t)$, we obtain $d\sigma(t)/dV(t)$ from the solution of Eq 4 . Alternatively, we can also use $\sigma$ vs. $V$ curve of Figure 3 to estimate $d\sigma/dV$. Comparison between the experimental and the predicted stress time plots is shown in Figure4 for all cycles. A good fit provides credence to our model. In Figure 5(d), we have also plotted velocity vs time during the slip phase for different values of $V_s$, These corresponds to the slip phases of first cycle of Figure 5 (a), the first and the second cycle of Figure 4 (b) and slip phase of Figure 5 (c). We see that, all velocities reach the asymptotic value $125\ \mu m.s^{-1}$ except for the case of the pulling velocity of $350\ \mu m.s^{-1}$ where the asymptotic value is $350\ \mu m.s^{-1}$ itself.

Next, we solve Eq 15 to estimate the variation of the attached chain fraction with time. The parameters used for simulation are $N_{c0} = 2.07 \times 10^{12}$, $u_c/\tau = 1.4 \times 10^{-4}\ s^{-1}$ and $\tau_a = 10\ s$ the parameter $\lambda M/kT = 2.47 \times 10^{12} m^{-2}$ is the same as that used for the slip phase calculation.

Figure 6 shows variation in the fraction of the attached chains with time. Sharp fall in the fraction of the attached chains is seen in all cases. The values of the stick time obtained from the simulation are $t_s = 18.3s$ for $V_0 = 50\ \mu m.s^{-1}$ and $t_s = 9.3s$ for $V_0 = 50\ \mu m.s^{-1}$. These are comparable with the experimental values of $18.1s$ and $t_s = 8.8\ s$ respectively.

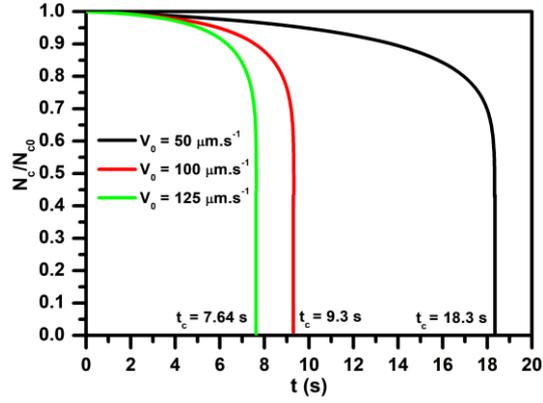

**Fig.6 : Variation with time of the fraction of the attached chains to the hard surface.**

Before concluding the discussion, importance of the stiffness of the block (i.e. $G_d/h$) needs to be emphasised. During the slip phase, when the sliding velocity falls below the pulling velocity $V_0$, pulling force begins to oppose the sticking force. If the stiffness of the block is large enough, then it can balance the sticking force before the sliding velocity reaches zero value. In such a case the sliding surface does not stick to the hard surface and the sliding dynamics may follow limit cycle. We have not included this case in the scope of our analysis.

## 4. Conclusions:

In the present work, we have analysed the stick-slip motion of a soft solid sliding on a hard surface. We have shown that during the stick phase, the sliding surface attains zero velocity. This observation is consistent with the sliding of elastomeric gelatine block on a glass surface. We have modeled the stick phase in terms of population balance of attached chains, where each chain is bonded to several sites on the surface. It is shown that slip occurs when the chains undergo debonding catastrophically. The slip phase is also modeled based on population balance of bonds which undergo aging at low sliding velocities. It is shown that the sticking occurs at a point when the frictional stress attains a minimum with respect to velocity. The model has been validated with the experimental data of Baumberger et al. (2002). This exercise has provided us with a greater insight into the dynamics of friction between a soft polymeric solid and a hard surface.